\documentclass[twocolumn,prl,superscriptaddress,showpacs,amssymb,amsmath,amsfonts,aps,floatfix]{revtex4-1}
\usepackage{graphicx}

\begin{document}
\title{Hard Two-body Photodisintegration of $^3{\rm{He}}$}

\newcommand*{\ANL}{Argonne National Laboratory, Argonne, Illinois 60439}
\newcommand*{\ANLindex}{1}
\affiliation{\ANL}
\newcommand*{\ASU}{Arizona State University, Tempe, Arizona 85287-1504}
\newcommand*{\ASUindex}{2}
\affiliation{\ASU}
\newcommand*{\UCLA}{University of California at Los Angeles, Los Angeles, California  90095-1547}
\newcommand*{\UCLAindex}{3}
\affiliation{\UCLA}
\newcommand*{\CANISIUS}{Canisius College, Buffalo, NY}
\newcommand*{\CANISIUSindex}{4}
\affiliation{\CANISIUS}
\newcommand*{\CMU}{Carnegie Mellon University, Pittsburgh, Pennsylvania 15213}
\newcommand*{\CMUindex}{5}
\affiliation{\CMU}
\newcommand*{\CUA}{Catholic University of America, Washington, D.C. 20064}
\newcommand*{\CUAindex}{6}
\affiliation{\CUA}
\newcommand*{\SACLAY}{CEA, Centre de Saclay, Irfu/Service de Physique Nucl\'eaire, 91191 Gif-sur-Yvette, France}
\newcommand*{\SACLAYindex}{7}
\affiliation{\SACLAY}
\newcommand*{\CNU}{Christopher Newport University, Newport News, Virginia 23606}
\newcommand*{\CNUindex}{8}
\affiliation{\CNU}
\newcommand*{\UCONN}{University of Connecticut, Storrs, Connecticut 06269}
\newcommand*{\UCONNindex}{9}
\affiliation{\UCONN}
\newcommand*{\HALIFAXDU}{Dalhousie University, Halifax, Nova Scotia B3H 3J5, Canada}
\newcommand*{\HALIFAXDUindex}{9}
\affiliation{\HALIFAXDU}
\newcommand*{\DUKE}{Duke University, Durham, North Carolina 27708}
\newcommand*{\DUKEindex}{9}
\affiliation{\DUKE}
\newcommand*{\EDINBURGH}{Edinburgh University, Edinburgh EH9 3JZ, United Kingdom}
\newcommand*{\EDINBURGHindex}{10}
\affiliation{\EDINBURGH}
\newcommand*{\FU}{Fairfield University, Fairfield CT 06824}
\newcommand*{\FUindex}{11}
\affiliation{\FU}
\newcommand*{\FIU}{Florida International University, Miami, Florida 33199}
\newcommand*{\FIUindex}{12}
\affiliation{\FIU}
\newcommand*{\FSU}{Florida State University, Tallahassee, Florida 32306}
\newcommand*{\FSUindex}{13}
\affiliation{\FSU}
\newcommand*{\Genova}{Universit$\grave{a}$ di Genova, 16146 Genova, Italy}
\newcommand*{\Genovaindex}{14}
\affiliation{\Genova}
\newcommand*{\GWUI}{The George Washington University, Washington, DC 20052}
\newcommand*{\GWUIIindex}{15}
\affiliation{\GWUI}
\newcommand*{\HUJI}{The Hebrew University of Jerusalem, 91904, Israel}
\newcommand*{\HUJIindex}{15}
\affiliation{\HUJI}
\newcommand*{\ISU}{Idaho State University, Pocatello, Idaho 83209}
\newcommand*{\ISUindex}{16}
\affiliation{\ISU}
\newcommand*{\ILLINOIS}{University of Illinois at Urbana-Champaign, Urbana, IL 61801}
\newcommand*{\ILLINOISindex}{16}
\affiliation{\ILLINOIS}
\newcommand*{\INFNFE}{INFN, Sezione di Ferrara, 44100 Ferrara, Italy}
\newcommand*{\INFNFEindex}{17}
\affiliation{\INFNFE}
\newcommand*{\INFNFR}{INFN, Laboratori Nazionali di Frascati, 00044 Frascati, Italy}
\newcommand*{\INFNFRindex}{18}
\affiliation{\INFNFR}
\newcommand*{\INFNGE}{INFN, Sezione di Genova, 16146 Genova, Italy}
\newcommand*{\INFNGEindex}{19}
\affiliation{\INFNGE}
\newcommand*{\INFNRO}{INFN, Sezione di Roma Tor Vergata, 00133 Rome, Italy}
\newcommand*{\INFNROindex}{20}
\affiliation{\INFNRO}
\newcommand*{\INFNTE}{INFN, Gruppo collegato Sanit\`a and Istituto Superiore di Sanit\`a, Department TESA, I-00161 Rome, Italy}
\newcommand*{\INFNTEindex}{21}
\affiliation{\INFNTE}
\newcommand*{\ORSAY}{Institut de Physique Nucl\'eaire ORSAY, Orsay, France}
\newcommand*{\ORSAYindex}{21}
\affiliation{\ORSAY}
\newcommand*{\ITEP}{Institute of Theoretical and Experimental Physics, Moscow, 117259, Russia}
\newcommand*{\ITEPindex}{22}
\affiliation{\ITEP}
\newcommand*{\JMU}{James Madison University, Harrisonburg, Virginia 22807}
\newcommand*{\JMUindex}{23}
\affiliation{\JMU}
\newcommand*{\KENT}{Kent State University, Kent, Ohio 44242}
\newcommand*{\KENTindex}{24}
\affiliation{\KENT}
\newcommand*{\KENTUCKY}{University of Kentucky, Lexington, Kentucky 40506}
\newcommand*{\KENTUCKYindex}{24}
\affiliation{\KENTUCKY}
\newcommand*{\KNU}{Kyungpook National University, Daegu 702-701, Republic of Korea}
\newcommand*{\KNUindex}{24}
\affiliation{\KNU}
\newcommand*{\LPSC}{LPSC, Universite Joseph Fourier, CNRS/IN2P3, INPG, Grenoble, France}

\newcommand*{\LPSCindex}{25}
\affiliation{\LPSC}
\newcommand*{\LBNL}{Lawrence Berkeley National Laboratory, Berkeley, California 94720}
\newcommand*{\LBNLindex}{26}
\affiliation{\LBNL}
\newcommand*{\MIT}{Massachusetts Institute of Technology, Cambridge, Massachusetts 02139}
\newcommand*{\MITindex}{26}
\affiliation{\MIT}
\newcommand*{\UNH}{University of New Hampshire, Durham, New Hampshire 03824-3568}
\newcommand*{\UNHindex}{26}
\affiliation{\UNH}
\newcommand*{\NRCN}{NRCN, P.O.Box 9001, Beer-Sheva 84190, Israel}
\newcommand*{\NRCNindex}{26}
\affiliation{\NRCN}
\newcommand*{\NSU}{Norfolk State University, Norfolk, Virginia 23504}
\newcommand*{\NSUindex}{27}
\affiliation{\NSU}
\newcommand*{\OHIOU}{Ohio University, Athens, Ohio  45701}
\newcommand*{\OHIOUindex}{28}
\affiliation{\OHIOU}
\newcommand*{\ODU}{Old Dominion University, Norfolk, Virginia 23529}
\newcommand*{\ODUindex}{29}
\affiliation{\ODU}
\newcommand*{\RPI}{Rensselaer Polytechnic Institute, Troy, New York 12180-3590}
\newcommand*{\RPIindex}{30}
\affiliation{\RPI}
\newcommand*{\ROMAII}{Universit$\grave{a}$ di Roma Tor Vergata, 00133 Rome Italy}
\newcommand*{\ROMAIIindex}{31}
\affiliation{\ROMAII}
\newcommand*{\RUTGERS}{Rutgers, The State University of New Jersey, Piscataway, New Jersey 08855}
\newcommand*{\RUTGERSindex}{32}
\affiliation{\RUTGERS}
\newcommand*{\HALIFAXSM}{Saint Mary's University, Halifax, Nova Scotia B3H 3C3, Canada}
\newcommand*{\HALIFAXSMindex}{33}
\affiliation{\HALIFAXSM}
\newcommand*{\MSU}{Skobeltsyn Nuclear Physics Institute, 119899 Moscow, Russia}
\newcommand*{\MSUindex}{32}
\affiliation{\MSU}
\newcommand*{\SCAROLINA}{University of South Carolina, Columbia, South Carolina 29208}
\newcommand*{\SCAROLINAindex}{33}
\affiliation{\SCAROLINA}
\newcommand*{\UTELAVIV}{Tel Aviv University, Tel~Aviv 69978, Israel}
\newcommand*{\UTELAVIVindex}{34}
\affiliation{\UTELAVIV}
\newcommand*{\UTEXAS}{The University of Texas at Austin, Austin, Texas 78712}
\newcommand*{\UTEXASindex}{35}
\affiliation{\UTEXAS}
\newcommand*{\TEMPLE}{Temple University, Philadelphia, Pennsylvania 19122}
\newcommand*{\TEMPLEindex}{35}
\affiliation{\TEMPLE}
\newcommand*{\JLAB}{Thomas Jefferson National Accelerator Facility, Newport News, Virginia 23606}
\newcommand*{\JLABindex}{34}
\affiliation{\JLAB}
\newcommand*{\UNIONC}{Union College, Schenectady, NY 12308}
\newcommand*{\UNIONCindex}{35}
\affiliation{\UNIONC}
\newcommand*{\UTFSM}{Universidad T\'{e}cnica Federico Santa Mar\'{i}a, Casilla 110-V Valpara\'{i}so, Chile}
\newcommand*{\UTFSMindex}{36}
\affiliation{\UTFSM}
\newcommand*{\GLASGOW}{University of Glasgow, Glasgow G12 8QQ, United Kingdom}
\newcommand*{\GLASGOWindex}{37}
\affiliation{\GLASGOW}
\newcommand*{\VIRGINIA}{University of Virginia, Charlottesville, Virginia 22901}
\newcommand*{\VIRGINIAindex}{38}
\affiliation{\VIRGINIA}
\newcommand*{\WM}{College of William and Mary, Williamsburg, Virginia 23187-8795}
\newcommand*{\WMindex}{39}
\affiliation{\WM}
\newcommand*{\YEREVAN}{Yerevan Physics Institute, 375036 Yerevan, Armenia}
\newcommand*{\YEREVANindex}{40}
\affiliation{\YEREVAN}

\newcommand*{\NOWMSU}{Skobeltsyn Nuclear Physics Institute, 119899 Moscow, Russia}
\newcommand*{\NOWORSAY}{Institut de Physique Nucl\'eaire ORSAY, Orsay, France}
\newcommand*{\NOWINFNGE}{INFN, Sezione di Genova, 16146 Genova, Italy}
\newcommand*{\NOWROMAII}{Universit$\grave{\textrm{a}}$ di Roma Tor Vergata, 00133 Rome Italy}
\newcommand*{\DEC}{deceased}
 %%%%%%%%%%%%%% END OF Latex Macros for institute addresses  %%%%%%%%%%%%%%%%%%%%%%%%% 

\author{I.~Pomerantz}
\affiliation{\UTEXAS}
\affiliation{\UTELAVIV}
\author{Y.~Ilieva}
\affiliation{\SCAROLINA}
\author{R.~Gilman}
\affiliation{\RUTGERS}
\affiliation{\JLAB}
\author{D.~W.~Higinbotham}
\affiliation{\JLAB}
\author{E.~Piasetzky}
\affiliation{\UTELAVIV}
\author{S.~Strauch}
\affiliation{\SCAROLINA}

\author {K.P. ~Adhikari} 
\affiliation{\ODU}
\author {M.~Aghasyan} 
\affiliation{\INFNFR}
\author{K.~Allada}
\affiliation{\KENTUCKY}
\author {M.J.~Amaryan} 
\affiliation{\ODU}
% ...

\collaboration{The CLAS and Hall-A Collaborations}
\noaffiliation

\begin{abstract}

We have measured cross sections for the $\gamma^3\textrm{He}\to pd$
reaction at photon energies of 0.4--1.4 GeV and a center-of-mass angle of 90$^\circ$. We observe dimensional scaling above 0.7 GeV at  this center-of-mass angle.  This is the first observation of dimensional scaling in the photodisintegration of a nucleus heavier than the deuteron. 
\end{abstract}


\maketitle

%% why as follows, does not agree with above
%\linenumbers


\label{sec:introduction}

%\begin{linenumbers}

Dimensional scaling laws
directly relate the energy dependence of the 
high-$t$ invariant cross sections to the number
of constituents of the hadrons involved in the process. 
The origin of dimensional scaling is the scale 
invariance of the interactions among hadron constituents, 
and, thus, it naturally reflects the property of asymptotic 
freedom of QCD at small distance scales. 
These laws  state that at fixed center-of-mass (c.m.)\
angle, the cross section of an exclusive two-body-to-two-body 
nuclear reaction at large $s$ (the total c.m. energy squared) and 
$t$ (the four-momentum transfer squared) is      
\begin{equation}
{d\sigma \over dt} \propto s^{2 - n_{i} - n_{f}} = s^{- n}
\end{equation}
where $n_{i}$ and $n_{f}$ are the total number of elementary fields in
the initial and final states that carry a finite fraction of particle 
momentum; \textit{e.g.}, 3 for a nucleon. 
Table~\ref{reactions_list} presents the %strong 
experimental evidence for the success of
these scaling laws.

\begin{table}[h]
\begin{ruledtabular}
\begin{tabular}{ccccccc}
  Reaction 				& $s$ 	& $\theta_{c.m.}$&$n$ & $n$ &Reference\\
					& GeV$^2$& deg. 	&Predicted	&Measured &	\\
\hline
  $pp \rightarrow pp$ 			& 15-60 & 38-90 & 10 	& 9.7$\pm$0.5 &\cite{PhysRevD.10.2300}\\
  $p\pi^- \rightarrow p\pi^-$ 		& 14-19 & 90    & 8 	& 8.3$\pm$0.3 &\cite{PhysRevD.21.2445}\\
  $\gamma p \rightarrow \gamma p$	& 7-12 & 70-120   &6 	& 8.2$\pm$0.5 &\cite{PhysRevLett.98.152001}\\
  $\gamma p \rightarrow \rho^0 p$	& 6-10 & 80-120   &7 	& 7.9$\pm$0.3 &\cite{battaglieri2001photoproduction}\\
  $\gamma p \rightarrow p \pi^0$	& 8-10  & 90    & 7 	& 7.6$\pm$0.7 &\cite{PhysRevC.60.052201}\\
  $\gamma p \rightarrow n\pi^+$ 	& 1-16  & 90    & 7 	& 7.3$\pm$0.4 &\cite{PhysRevLett.91.022003}\\
  $\gamma p \rightarrow K^+ \Lambda$  & 5-8 & 84-120 & 7 & 7.1$\pm$0.1& \cite{PhysRevC.83.025207}\\
  $\gamma d \rightarrow pn$ 		& 1-4   & 50-90 &11 	& 11.1$\pm$0.3 &\cite{Napolitano:1988uu,Freedman:1993nt,Belz:1995ge,Bochna:1998ca,Schulte:2001se,Schulte:2002tx,Mirazita:2004rb,Rossi:2004qm}\\
  $\gamma pp \rightarrow pp$ 		& 2-5   & 90    & 11 	& 11.1$\pm$0.1 &\cite{Pomerantz2010106}\\
  $\gamma ^3He \rightarrow pd$ 		&11-15.5&90	&17	&17.0$\pm$ 0.6&(this work)
\end{tabular}
\caption{
\label{reactions_list}Selected hard exclusive hadronic and nuclear reactions that have been previously measured.}
\end{ruledtabular}
\end{table}

Dimensional scaling is well founded and expected at asymptotic energies,
where the available energy in the c.m. is  much higher than the
mass of the system. Under these circumstances, the only scale available is
the energy and the $s$ dependence arises from the norm of the active fields.
However, for some reactions, the data show
evidence for dimensional scaling even when $s$ is roughly equal to the squared mass of the system, as is the case reported here.

To date there is no common model or theory that can describe
all the data listed in Table~\ref{reactions_list} in a consistent
manner. 
For processes on nuclear targets, phenomenological extensions 
of $p$QCD based on factorization~\cite{Brodsky:1983kb, Frankfurt:1999ik, Sargsian:2008zm} have been developed and have 
shown limited success. 
A common feature of model interpretations of dimensional 
scaling, such as~\cite{Matveev:1973ra, Brodsky:1973kr, PhysRevLett.90.241601}, is the dominance of a hard scattering mechanism in the 
reaction dynamics. It was, however, also discussed
that soft QCD processes~\cite{Radyushkin:2006} or destructive interference among resonances~\cite{Zhao:2003} can 
mimic scaling at medium energies.  

From the 
standpoint of a non-perturbative approach, the scaling laws have been reviewed and derived 
using the AdS/CFT correspondence 
between string theories in Anti-de Sitter space-time and 
conformal field theories in physical space-time~\cite{Polchinski:2001tt}. 
Within this approach, the interactions between hadron 
constituents are scale-invariant at very short but also 
at very large distances in the so-called 
``conformal window" where the effective strong coupling 
is large but constant, \textit{i.e.}, scale independent. 
Thus, dimensional scaling laws may be probing in fact the 
limits of two very different dynamical regimes of asymptotically 
large $t$ and $s$, and of small $t$. 
In order to understand better the origin of scaling, we would need 
to also probe rigorously exclusive nuclear processes at very small
$t$. 

For reactions that are dominated by resonances, the study of 
scaling at smaller $t$ is difficult since the resonances make it hard to determine
whether scaling is being observed. 
We chose to probe dimensional scaling in the reaction 
$\gamma ^3{\rm He} \to pd$ in the photon energy range 0.4 -- 1.4 GeV.
In this energy range, photoreactions on the proton and deuteron
have shown signatures of scaling
\cite{PhysRevLett.91.022003,Napolitano:1988uu,Freedman:1993nt,Belz:1995ge,Bochna:1998ca,Schulte:2001se,Schulte:2002tx,Mirazita:2004rb,Rossi:2004qm},
but their interpretation is unclear. 
This reaction has the advantage that resonance mechanisms 
are suppressed (as shown by low-energy studies)~\cite{Picozza:1970npa}. 
In addition, there is evidence that two- and three-body 
mechanisms are important at large c.m. angles, \textit{i.e.},
the momentum transfer is shared among two or three nucleons 
so that the average momentum transfer to each quark 
constituent would be small (maybe in the range of 
the ``conformal window").
No other study of dimensional scaling in an $A$ = 3 system 
has been done before.
Moreover, our measurement is the first of this reaction in the 
GeV energy region.
As previous measurements of lepton-induced reactions have only involved $A$ = 1 or 2, 
the expected quark-counting scaling power of 
$d\sigma/dt \propto s^{-17}$ is higher than any previous
observation in 
%leptoproduction. ???
photo- or hadronic reactions.

\label{sec:experiment}
The data presented here were taken as part of Jefferson lab (JLab) Experiments 
03-101 and 93-044, which ran at the continuous electron beam accelerator facility (CEBAF) 
 in Hall A~\cite{Alcorn:2004sb} and in Hall B~\cite{Berman:1993pr}, respectively. 

E03-101 was a measurement of the $\theta_{c.m.}$ = 90$^{\circ}$ 
energy dependence of the  
$^3$He($\gamma,pp)n_{spectator}$ reaction~\cite{Pomerantz2010106}. 
At an incident electron energy of 1.656 GeV we also took data at two
kinematical settings which did not match the $\theta_{c.m.}$ =
90$^{\circ}$ condition for that reaction. 
In these two kinematics we could identify two-body photodisintegration 
of the $^3$He into a proton and a deuteron at angles corresponding to 
$\theta_{p\ c.m.}$ = 85$^{\circ}$.

\begin{figure}
\begin{center}
\includegraphics[width=\linewidth]{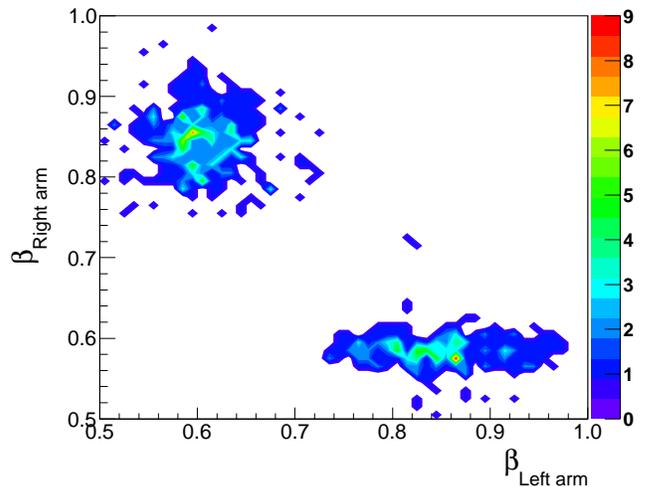}
\caption{\label{fig:betas} (Color online) 
The $\beta$ distribution of particles detected in coincidence by the two High Resolution Spectrometers in Hall A/E03-101.
The widths of the peaks result from 
the calibration and time resolution of the scintillators, 
from the momentum acceptance of the spectrometers 
($\Delta p \sim \pm$3.5\%), 
and the uncertainty of the path-length correction.
The different scintillators in the two spectrometers lead to different widths of the distributions.}
\end{center}
\end{figure}

In this experiment, untagged bremsstrahlung photons were generated 
when the electron beam of 1.656 GeV impinged on a copper radiator. 
The 6\%-radiation-length radiator was located in the scattering
chamber 38~cm upstream of the center of a 20-cm long cylindrical 
0.079\ g/cm$^3$ ${\rm ^3He}$ gas target.
The size of the photon beam spot on the target, $\sim$2 mm, results 
from electron beam rastering intended to distribute the heat load 
across the target.
The size of the target is much smaller than the $\sim$1-cm size of the target windows 
and apertures.   
Protons and deuterons from the target were detected in coincidence 
with the Hall-A high-resolution spectrometers (HRSs) \cite{Alcorn:2004sb}. 
The two spectrometers were set symmetrically on the two sides of the 
beam line in two kinematical settings corresponding to 
central momenta of 1.421 GeV/$c$ at a scattering angle of 63.16$^{\circ}$ 
and 1.389  GeV/$c$ at a scattering angle of 65.82$^{\circ}$.  

For each spectrometer, the scattering angles, momenta, and interaction 
positions at the target were reconstructed from trajectories measured 
with vertical drift chambers (VDCs) located in the focal plane.  
Two planes of plastic scintillators provided triggering and
time-of-flight information for particle identification. 
Figure~\ref{fig:betas} shows the speed, $\beta$, of the two 
particles detected in coincidence.
One clearly sees protons and deuterons in coincidence, with no 
visible backgrounds, such as $pp$ and $dd$ coincidences, or pions. 

In analyzing the data from E03-101, 
the incident photon energy of the untagged beam
was reconstructed event by event from the momentum and angles of the 
scattered particles under the assumption of two-body $pd$ final-state kinematics. 
In order to assure the validity of this assumption and reduce backgrounds, 
the analysis is limited to events that fulfill two energy and momentum constraints:
\begin{enumerate}
	\item $p_{T \, missing} \equiv p_{T(p)}+p_{T(d)}$ $<$ 5 MeV/$c$, and
%	\item $E_{missing}=E_d+E_p-M_{^3He}-P_{pz}-P_{dz}$ $<$ 5 MeV. 
	\item $\alpha_{missing}\equiv
           \alpha_d+\alpha_p-\alpha_{^3He}-\alpha_\gamma$ $<$ $5\cdot10^{-3}$. 
\end{enumerate}
where $\alpha$ is the light cone variable for each particle participating in the reaction:
\begin{equation}
\alpha_X = 
A{{E^X-p^X_z}\over{E^A-p^A_z}} \approx  {E_X-p^X_z \over m_A / A},
\end{equation}
where $A=3$ is the mass number, $E^X$ and $p^X$ are respectively the energy and momentum of the particle,
$m_A$, $E^A$ and $p^A$ are the nucleus mass, energy and momentum respectively,
and the $z$ direction is the direction of the incident 
photon beam. With the above definitions, $\alpha_\gamma$ is 
zero, while $\alpha_{^3\rm{He}}$ is 3.

Simulation of pion production with this process shows that with 
these cuts the contamination of non 2-body events is negligible.

The detected proton-deuteron pairs can result from either a photon 
or an electron disintegrating the ${\rm ^3He}$ nucleus. 
We took data with the radiator in and out of the beam, to extract 
the number of events resulting from photons produced in the 
bremsstrahlung radiator ~\cite{Schulte:2001se,Pomerantz2010106}.
Event selection cuts on the target vertex and coincidence between the 
two spectrometers were applied using the same techniques as \cite{Pomerantz2010106}.
The finite acceptance correction was determined using the standard Hall-A Monte-Carlo 
simulation software MCEEP~\cite{MCEEP}.

The sources for the systematic uncertainties for E03-101 are described in \cite{Pomerantz2010106};
for this analysis they are dominated by the finite acceptance correction, which is at the 4\% -- 11\% level.

Experiment E93-044 used the 
CEBAF large acceptance spectrometer (CLAS) to
measure various photoproduction reactions on $^3\textrm{He}$ and 
$^4\textrm{He}$ targets. 
A collimated, tagged, real-photon beam was produced using the 
bremsstrahlung tagging facility in Hall B \cite{Sober2000263}.
Photons with energies between 0.35 and 1.55 GeV were incident on 
a 18-cm long cryogenic liquid $^3$He target positioned in the center of the
CLAS. 
The outgoing protons and deuterons were tracked in the six toroidal 
magnetic spectrometers (sectors) of the CLAS. 
Their trajectories were measured by three layers
of drift chambers surrounding the target. 
%The particles' 
Particle time of flight was measured by $6\times 57$
scintillators (TOF) enclosing the CLAS outside of the magnetic field. 
The CLAS covers a polar angular range from $8^\circ$ to $142^\circ$ 
and an azimuthal angular range from $0^\circ$ to $360^\circ$, excluding 
the angles where the torus coils are located.
More details about the CLAS and experiment E93-044 
can be found in~\cite{Mecking:2003zu} and \cite{Niccolai:2004prc}, respectively.

In the analysis of data from E93-044, 
protons and deuterons were identified from 
momentum and time-of-flight measurements. 
Only events with one proton and one deuteron originating from the target
were used for further analysis.
Accidental and physics backgrounds were reduced by 
applying kinematic cuts making use of the constraints provided by 
two-body kinematics when both final-state particles are detected. 
A detailed discussion of the selection cuts can be found
in~\cite{Ilieva:note}. 

\begin{figure}
\begin{center}
\includegraphics[width=\linewidth,height=180px,trim=0.5cm 1cm 2cm 2cm]{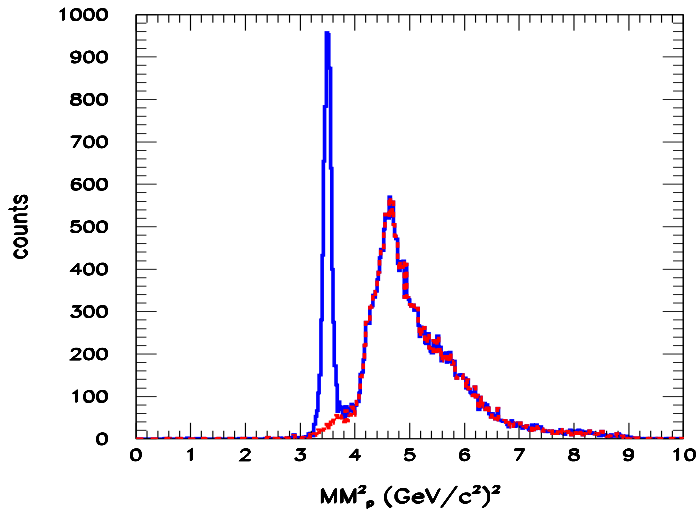}
\caption{\label{fig:clasanalysis} (Color online)
Event distributions measured by CLAS, of the missing-mass of the proton, $MM^2_p$ for 
proton c.m. angle of $90^\circ$, with (dashed red line) and without 
(blue solid line) the kinematic cuts, are shown. 
Events from the $pd$ final state are clearly identified in the peak. 
Accidental and multipion events give rise to the background. 
The background distribution (events rejected by the kinematic cuts) exhibits smooth behavior under the 
deuteron peak and reproduces nicely the background
shape outside of the peak.}
\end{center}
\end{figure}

Figure~\ref{fig:clasanalysis} demonstrates the effect of the kinematic
cuts on the proton missing-mass-squared, $MM_p^2$, distribution at a proton
c.m. angle of $90^\circ$. The proton missing-mass-squared is calculated as 
$MM_p^2=(\tilde{p}_\gamma+\tilde{p}_{^3\textrm{He}}-\tilde{p}_p)^2$, where
$\tilde{p}_\gamma$, $\tilde{p}_{^3\textrm{He}}$, and $\tilde{p}_p$ are the four-momentum
vectors of the beam, target, and detected final-state proton, respectively.
The initial event distribution, before our kinematic cuts are applied, shows
a well pronounced peak at around 3.5 (GeV/$c^2$)$^2$ (which corresponds to the square of the
%nominal 
deuteron mass), followed by a broader structure above 3.8 (GeV/$c^2$)$^2$. While
the peak contains predominantly the $pd$ events of interest, the broader structure contains 
background events
produced in the reaction $\gamma ^3\textrm{He}\to pdX$, where $X$ could be one or more
missing particles. One could see that the low-mass tail of the background events extends under the $pd$ peak.
Our kinematic cuts select the good $pd$ events from the initial sample and reject background events.   
For simplicity, in Fig.~\ref{fig:clasanalysis} we show the events rejected by our kinematic cuts overlaid with the initial distribution.
These background events 
exhibit smooth behavior under the 
deuteron peak and reproduce the background shape 
outside of the peak, as expected.
The uncertainty of the yield extraction due to 
the remaining background events is $(2.30\pm 0.63)\%$. 

The CLAS acceptance for the reaction $\gamma ^3\textrm{He}\to pd$ 
was evaluated by generating $2\times 10^7$ phase-space events and processing 
them through GSIM~\cite{gsim}, a GEANT-3 program that simulates the CLAS. 
The CLAS acceptance for $pd$ events at a c.m. angle of $90^\circ$ is $\sim 71\%$. 
The systematic uncertainty of the acceptance is $<10\%$. 
The photon flux was calculated using the standard CLAS software~\cite{gflux} and has
an uncertainty of 4.5$\%$~\cite{ilieva:pi0d}.
The uncertainty of the target length and density was 2$\%$ \cite{Niccolai:2004prc}. 
The total systematic uncertainty of the CLAS cross sections is $<11.4\%$,
with the CLAS acceptance being the dominant source.
The statistical uncertainties range from 2$\%$ to $40\%$ depending on the energy bin. 
Full details about the analysis of the CLAS data can be found in \cite{Ilieva:note}. 

\begin{figure}
\begin{center}
\includegraphics[width=\linewidth]{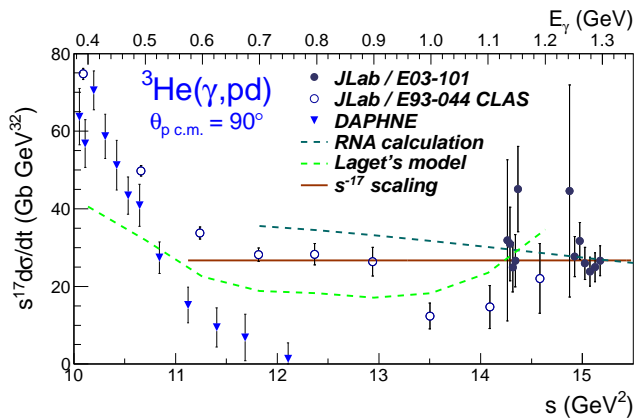}
\caption{The invariant cross section $d\sigma/dt$ multiplied
by $s^{17}$ to remove the expected energy dependence.
The DAPHNE data is taken from \cite{Isbert:1993xf}. 
The RNA ~\cite{Brodsky:1983kb}  calculation is normalized to our highest energy data point from JLab/E03-101.
The prediction of Laget is based on a diagrammatic hadronic model~\cite{Laget:1988hv}.
For a comment on the comparison between the CLAS and DAPHNE data see~\cite{CLAS:DAPHNE}.}
\label{fig:s17dsdt}
\end{center}
\end{figure}

Figure~\ref{fig:s17dsdt} shows the resulting cross sections from CLAS and Hall A compared 
to previously published data \cite{Isbert:1993xf} for $s$ $>$10 GeV$^2$. 
In the range of $s$ = 11.5 -- 15 GeV$^2$, the cross section falls by 
two orders of magnitude.
The falloff of our Hall-A and CLAS data is fit as $s^{-17 \pm 0.6}$, which is consistent
with the expected scaling degree of $n$ = $17$.
This is the first observation of high-energy cross-section
scaling for photodisintegration of an $A > 2$ system. 
We note that our data point at $s\sim 13.5$ GeV$^2$ is about 3.5 standard deviations
below the scaling prediction. Due to the limited statistics we have in this kinematic bin 
we cannot study in further detail whether the origin of this deviation is random or is due
to physics. 

Starting at threshold, the scaled invariant cross section, $s^{17} d\sigma/dt$, decreases 
smoothly to $E_\gamma$=0.7~GeV where it levels out, a transition different from 
meson photoproduction \cite{PhysRevLett.91.022003} or 
$pp$ breakup \cite{Pomerantz2010106}, 
where ``resonance-like" structures are observed.
Since our data are taken in the resonance region (at $W < 2$ GeV, where $W$ is the total
center-of-mass energy in the initial state assuming a free nucleon target), 
this suggests that two- and three-nucleon mechanisms dominate
 the reaction dynamics or nucleon resonance contributions are strongly suppressed. 

The scaled cross section of $\sim$30 Gb$\cdot$GeV$^{32}$ corresponds to an
invariant cross section of $d\sigma/dt$ $\sim$ 0.4 nb/GeV$^2$ for $E_{\gamma}$ $\sim$ 1.3 GeV.
The corresponding cross section for $\gamma d \to pn$ at this energy is
about 30 nb/GeV$^2$, about two orders of magnitude larger,
while the scaled cross section for $\gamma ^3{\rm He} \to pp + n_{spectator}$ 
at this energy is about 13 nb/GeV$^2$, about 30 times larger.
If one adopts the view that large momentum transfer reactions select initial
states in which all the quarks and nucleons are close together, it is much
more likely that there is a short-range, and thus high-momentum, $pn$ pair
than $pp$ pair. This is what has been found in recent studies for nucleons
above the Fermi surface that have momenta of several hundred 
MeV/$c$ \cite{Subedi:2008zz,Shneor:2007tu}.
Furthermore, in $^3$He there is nearly as large a probability for a 
short-range $pd$ pair as for a $pp$ pair~\cite{Sargsian:2005ru}. 

The reduced nuclear amplitudes (RNA)
prescription~\cite{Brodsky:1983kb} 
was developed as a way of extending the applicability of $p$QCD to 
lower energy and momentum scales, by factoring out
non-perturbative dynamics related to hadron structure 
through phenomenologically determined hadronic form factors.
It should be noted that deuteron photodisintegration follows the 
dimensional scaling better than it follows the RNA prediction~\cite{Mirazita:2004rb}.
The RNA prescription for $\gamma ^3{\rm He} \to p d$ is:
\begin{equation}
{{d\sigma}\over{dt}} \propto {{1}\over{(s-m_{^3{\rm He}}^2)^2}}
 F^2_p(\hat{t}_p) F^2_d(\hat{t}_d) {{1}\over{p^2_T}} f^2(\theta_{c.m.}).
\end{equation}
Here $F_p$ ($F_d$) is the proton (deuteron) form factor, 
$\hat{t}_p$ ($\hat{t}_d$) is the momentum transfer to the proton (deuteron) 
and $f$ is an unknown function of the c.m. angle that must be
determined from experimental data.
The overall normalization is also unknown, and ideally should be 
determined from data at asymptotically large momentum transfer.
Figure~\ref{fig:s17dsdt} shows the RNA prediction, 
normalized to our highest energy data point from E03-101. The 
experimental data appear to agree better with dimensional scaling 
than with the RNA prediction.

If the reaction dynamics 
are dominated by nucleon rescattering, it appears that 
hard $pn$ rescattering is more likely than hard $pp$ rescattering -- 
which is known to be the case from cancellations in the $pp$ amplitude 
\cite{Sargsian:2008zm}, 
and that hard $pd$ rescattering is also suppressed, due 
to the likelihood of breaking up the deuteron in a hard scattering 
and the small probability of a pickup reaction that create a deuteron 
from a scattered proton or neutron.
Calculation of the cross section in the framework of the Hard 
Rescattering Model (HRM) \cite{Frankfurt:1999ik} using  
elastic $pd$ scattering data is in preparation \cite{priv:misak}. 

The model of Laget~\cite{Laget:1988hv} is a hadronic model based on a diagrammatic approach 
for the calculation of the dominant one-, two-, and three-body mechanisms contributing
to the reaction. This calculation provides good accounting of the absolute magnitude of 
the cross section and reproduces the scaling exhibited by the data over a limited energy range.
Overall, the data appear to agree better with dimensional scaling
than with the model.    

We observe the onset of scaling at $\theta_{c.m.}$ = $90^\circ$ at a momentum transfer to the deuteron $|t|>0.64$ (GeV/$c$)$^2$ and a transverse momentum $p_\perp > 0.95$ GeV/$c$.
These momentum thresholds for scaling are remarkably low.
For other processes, such as deuteron photodisintegration, the onset 
of scaling has been observed at $p_\perp > 1.1$ GeV/$c$
\cite{Napolitano:1988uu,Freedman:1993nt,Belz:1995ge,Bochna:1998ca,Schulte:2001se,Schulte:2002tx,Mirazita:2004rb,Rossi:2004qm}.
Both the deuteron form factor and the reduced deuteron form factor~\cite{Brodsky:1983kb} 
show scaling at $|t| > 2$ (GeV/$c$)$^2$.
This comparison suggests that non-perturbative interpretation of our 
data may be more appropriate. 
Such interpretation in the framework of AdS/CFT means that the 
observed scaling is due to the near-constancy of the effective QCD 
coupling at low $Q$ (``conformal window'' \cite{PhysRevD.77.056007}) 
and we are in the non-perturbative regime of QCD. 
A further test of this interpretation would require data for this
process over a higher-energy range where the transition 
from non-perturbative to perturbative dynamics would manifest itself 
in breaking dimensional scaling. 
The latter would be observed again at asymptotically large invariants 
when $p$QCD sets in.

Our result not only indicates that dimensional scaling is a feature of 
non-perturbative strong dynamics, but also that QCD studies of 
nuclei are meaningful at energies as low as E$_\gamma$ = 0.7 GeV and that the three-nucleon 
bound system may be an equally good laboratory for such studies as the deuteron. 
Moreover, since the cross section for our process had been previously measured 
down to beam energies of a few MeV, our data combined with the low-energy data 
allow to map for the first time the transition from meson-nucleon to partonic degrees 
of freedom cleanly, without the complication of resonance structures, as has been the 
case in previous studies involving $A$ = 1 or $A$=2 nuclear systems.

\label{sec:summmary}

We have observed for the first time scaling in an exclusive reaction 
initiated by a photon beam and involving an $A$ = 3 nucleus, the two-body photodisintegration of $^3{\rm He}$.
The scaling power of $s^{-17}$ for $E_\gamma > 0.7$ GeV, is the
highest quark-counting power-law dependence observed to date in leptoproduction.
This is only one of a few examples of scaling in a nuclear photoreaction,
and it remains unclear whether there is any unified explanation for 
the various onsets of scaling in the different photoreactions.
If AdS/CFT correspondence is the proper framework to understand 
the origin of dimensional scaling, then the observed scaling is a
result of the near-constancy of the QCD coupling. 
This assumption may be validated through the study of this reaction 
in a higher energy range.

%\end{linenumbers}


%% References


%\begin{acknowledgements}

We thank S. J. Brodsky, L. L. Frankfurt, M. M. Sargsian and M. Strikman for helpful discussions.
We thank the JLab physics and accelerator divisions for their support.
This work was supported in part by the U.S.\ National Science Foundation 
under grant PHY-0856010, the U.S.\ Department of Energy,
the Israel Science Foundation, the US-Israeli Bi-National Scientific
Foundation, the Chilean Comisi\'on Nacional de Investigaci\'on Cient\'ifica y Tecnol\'ogica (CONICYT),
the Istituto Nazionale di Fisica Nucleare,
the French Centre National de la Recherche Scientifique,
the French Commissariat \`{a} l'Energie Atomique,
the UK Science and Technology Facilities Council (STFC),
the Scottish Universities Physics Alliance (SUPA),
and the National Research Foundation of Korea. Jefferson Science Associates operates
the Thomas Jefferson National Accelerator Facility under DOE
contract DE-AC05-06OR23177.
%\end{acknowledgements}
\bibliographystyle{apsrev4-1}
\bibliography{gammapd-refs}

\begin{thebibliography}{44}
\bibitem [{CLA()}]{CLAS:DAPHNE}%
  \BibitemOpen
  \href@noop {} {}\bibinfo {note} {At this time, the disagreement between the
  CLAS and DAPHNE data at $E_\gamma >$ 0.5 GeV is not well understood. The
  DAPHNE yields were corrected by a factor of up to 42$\%$ depending on photon
  energy and c.m. angle. Their analysis is more reliable at kinematics where
  the final state particles stop in the detector. The largest disagreement
  between the two data sets is at mid c.m. angles where both the proton and the
  deuteron have large momenta and do not stop in the DAPHNE detector. This may
  lead to a large systematic uncertainty in their correction
  factor.}\BibitemShut {Stop}%
\end{thebibliography}

\end{document}